\documentclass[preprint,preprintnumbers,amsmath,amssymb]{revtex4}
\usepackage{graphicx}
\begin{document}

\title{Geometrical origin of entropy during inflation from the STM theory of gravity}
\author{ $^{1}$ Jes\'us Romero and $^{1,2}$ Mauricio
Bellini \footnote{E-mail address: mbellini@mdp.edu.ar}}

\address{$^{1}$ Departamento de F\'{\i}sica, Facultad de Ciencias Exactas y
Naturales, Universidad Nacional de Mar del Plata, Funes 3350,
(7600) Mar del Plata,
Argentina.\\
$^2$ Consejo Nacional de Investigaciones Cient\'{\i}ficas y
T\'ecnicas (CONICET). }

\begin{abstract}
Using a recently introduced 5D Riemann flat metric, we investigate
the possibility of introducing dissipation in the dynamics of the
inflaton field on an effective 4D FRW metric, in the framework of
the STM theory of gravity.
\end{abstract}


\maketitle
\section{Introduction}

A number of higher-dimensional theories such as 11D
supergravity\cite{sup} and 10D superstrings\cite{sups} appeared
inspired by the 5D Kaluza-Klein (KK) theory\cite{kk} in the
sixties and seventies, all of them aiming at a general scheme of
unification\cite{1}. More recently, the so-called braneworld
scenario has emerged, according to which our 4D spacetime is
viewed as a hypersurface (the brane) isometrically embedded in an
5D space (the bulk)\cite{2}. The original version of the KK theory
assures, as a postulate, that the fifth dimension is compact. A
few years ago, a non-compactified approach to KK gravity, known as
Space-Time-Matter (STM) theory was proposed by Wesson and
collaborators\cite{3}. In this theory all classical physical
quantities, such as matter density and pressure, are susceptible
of a geometrical interpretation. Wesson's proposal also assumes
that the fundamental 5D space in which our usual spacetime is
embedded, should be a solution of the classical 5D vacuum Einstein
equations: $R_{AB}=0$. The mathematical basis of it is the
Campbell's theorem\cite{campbell}, which ensures an embedding of
4D general relativity with sources in a 5D theory whose field
equations are apparently empty. That is, the Einstein equations
$G_{\alpha\beta}=-8\pi G\,T_{\alpha\beta}$ (we use $c=\hbar=1$
units), are embedded perfectly in the Ricci-flat equations
$R_{AB}=0$. In simple terms, Wesson uses the fifth dimension to
model matter. An alternative version of 5D gravity, which is
mathematically similar, is the membrane theory. In this, gravity
propagates freely in 5D, into the "bulk", but the interactions of
particles are confined to a 4D hypersurface called "brane". Both
versions of 5D general relativity are in agreement with
observations. Inflationary cosmology is one of the most reliable
concepts in modern cosmology. In particular, the warm inflation
scenario\cite{warm} takes into account separately, the matter and
radiation energy fluctuations. In this scenario, the interaction
between the inflaton field and the particles of a thermal bath
with temperature $T$ produced during inflation, provides
slow-rolling of the inflaton towards the minimum of the potential
$V(\varphi)$. Hence, in this model slow-roll conditions are
physically well justified. The decay width ($\Gamma$) of the
produced particles grows with time, so when the inflaton
approaches the minimum of the potential there is no oscillation
around the minimum energetic configuration. This is due to
dissipation being too large with respect to the Hubble parameter
($\Gamma \gg H$).

The aim of this letter consists to explain how dissipative
dynamics of the inflaton field on a 4D Friedmann-Robertson-Walker
(FRW) metric, can be geometrically induced from a dynamically
foliated 5D Riemann flat metric, on which we shall define our
physical vacuum. Moreover, we are interested to study how this
dissipation is related to the fifth coordinate and the growth of
entropy, when the equivalence principle is broken.

\section{5D vacuum and field dynamics}

We consider the recently introduced 5D line element \cite{MET}
\begin{equation}\label{a1}
dS^{2}=\psi^{2}\,\frac{\Lambda(t)}{3}\,dt^{2}-\psi ^{2} \,e^{2\int
\sqrt{\Lambda(t)/3}\,dt}dr^{2}-d\psi ^{2},
\end{equation}
where $\Lambda(t)$ is a time-dependent function,
$dr^{2}=dx^2+dy^2+dz^2$ is the 3D Euclidean metric, $t$ is the
cosmic time and $\psi$ is the space-like extra dimension. Choosing
a natural unit system the function $\Lambda (t)$ has units of
$[length]^{-2}$. The metric (\ref{a1}) is Riemann flat and hence
is suitable to describe a 5D vacuum state in the context of
Space-Time-Matter (STM) theory of gravity (the reader can see
\cite{STM} and references therein).

To describe the system in an apparent vacuum, we shall consider
the action
\begin{equation}
^{(5)}I = {\Large\int} d^4 x \  d\psi \sqrt{\left|\frac{^{(5)}
 g}{^{(5)} g_0}\right|} \left(
\frac{^{(5)} R}{16\pi G}+ \frac{1}{2}g^{AB} \varphi_{,A}
\varphi_{,B} \right),
\end{equation}
where $^{(5)}g$ is the determinant of the covariant metric tensor
$g_{AB}$ ($A$ and $B$ run from $0$ to $4$):
\begin{equation}\label{ricci5}
^{(5)}g= \psi^8 \,\left(\frac{\Lambda}{3}\right) \, e^{6\int
\sqrt{\frac{\Lambda}{3}} dt}.
\end{equation}
For simplicity we shall consider $\varphi$ as a classical scalar
field, which is minimally coupled to gravity. The Lagrangian is
purely kinetic because we describe an apparent vacuum in absence
of interactions. In this sense, $\varphi$ can be considered as a
massless test field on the 5D vacuum. However, a more consistent
approach would take into account the quantum nature of $\varphi$.
The equation of motion for the field $\varphi$ is
\begin{equation}
\ddot\varphi + \left[3\sqrt{\frac{\Lambda}{3}} -
\frac{\dot\Lambda}{2\Lambda}\right] \dot\varphi -
\frac{\Lambda}{3} e^{-2\int \sqrt{\frac{\Lambda}{3}} dt}
\nabla^2\varphi  -  \frac{\Lambda}{3}\left[4\psi
\frac{\partial\varphi}{\partial\psi} + \psi^2
\frac{\partial^2\varphi}{\partial\psi^2}\right] =0,\label{dis}
\end{equation}
which describes the dynamics of $\varphi$ on the 5D vacuum
(\ref{a1}).

\section{Dissipative dynamics of $\varphi$ on a FRW metric}

We shall assume that the 5D spacetime (\ref{a1}) is dynamically
foliated on the fifth coordinate: $\psi=\psi(t)$, such that the
effective 4D hypersurface is
\begin{eqnarray}
&& \left.dS^2\right|_{\psi=\psi(t)}=
\left.d\sigma^2\right|_{\psi=\psi(t)} -\dot\psi^2(t) dt^2
\nonumber \\
&& \equiv \left[\psi^2(t)\frac{\Lambda(t)}{3} -
\dot\psi^2(t)\right]dt^2 - \psi^2(t) \, e^{2\int
\sqrt{\frac{\Lambda(t)}{3}} dt} \, dr^2, \label{u}
\end{eqnarray}
with the condition $\psi^2(t) {\Lambda(t)\over 3} - \dot\psi^2(t)
>0$, such that $g_{AB} U^A U^B =1$ [$U^A = {dX^A\over dS}$ are the
components of the penta-velocity]. However, we are interested to
study the evolution of $\varphi$ on the hypersurface
$\left.d\sigma^2\right|_{\psi=\psi(t)}$, which can be considered
as a brane on (\ref{u})
\begin{equation}\label{uu}
\left.d\sigma^2\right|_{\psi=\psi(t)} = g_{\mu\nu}\left(t,\vec
r,\psi(t)\right) \, dx^{\mu} dx^{\nu} \neq
\left.dS^2\right|_{\psi=\psi(t)}.
\end{equation}
Notice that the equivalence principle is broken on (\ref{uu}):
$g_{AB} U^A U^B \neq 1$. We can define $\rho=\rho_0+\Delta\rho$
and $p=p_0+\Delta p$, the energy density and the pressure on
$\left.d\sigma^2\right|_{\psi=\psi(t)}$ [described by (\ref{uu})]
in (\ref{u}): such that $\rho_0(t)$ and $p_0(t)$ are the energy
density and the pressure on the effective 4D hypersurface
$\left.dS^2\right|_{\psi=\psi(t)}$ in (\ref{u}). Furthermore, the
system describes an adiabatic expansion on (\ref{u}), so that
(from the thermodynamical point of view) it can be considered as a
closed system
\begin{equation}
\frac{d}{dt}\left[\rho_0\,a^3(t)\right] + p_0\,
\frac{d}{dt}\left[a^3(t)\right]=0.
\end{equation}
Hence, $\Delta\rho$ and $\Delta p$ comply with the following
equation on (\ref{uu}):
\begin{equation}\label{del}
\Delta\dot\rho+3 H \left(\Delta\rho+\Delta p\right) =
\frac{T}{a^3(t)} \,\dot{\cal S}.
\end{equation}
Here, $a(t)=\psi(t)\,e^{\int \sqrt{{\Lambda\over 3}}\,dt}$,
$H(t)=\dot a/a$, $T$ is the temperature of the system on
$\left.d\sigma^2\right|_{\psi=\psi(t)}$ and ${\cal S}$ is its
entropy. We can define $\gamma = 1+{\Delta p\over \Delta\rho}$, so
that the eq. (\ref{del}) can be rewritten as
\begin{equation}
\Delta\dot\rho+3 H \,\gamma \,\Delta\rho
=\frac{T}{a^3(t)}\dot{\cal S}.
\end{equation}
A very interesting case, of particular interest is
${\Delta p\over \Delta\rho}=1/3$. In this particular case we can
identify $\Delta\rho$ and $\Delta p$ with the radiation energy
density and its pressure: $\Delta\rho \equiv \rho_r$ and $\Delta p
\equiv p_r$. In this case $\gamma= 4/3$ and the system radiates
\begin{equation}\label{en}
\dot\rho_r+4 H\,\rho_r =\delta,
\end{equation}
where the interaction $\delta\equiv\frac{T}{a^3(t)} \dot{\cal S}
> 0$ is related with the variation of entropy ${\cal S}$. Notice
that the entropy ${\cal S}$ increases with time, so that both
sides in (\ref{en}) are positives.

The Lagrange equation describes a non-conservative system:
\begin{equation}
\frac{\partial ^{(4)}{\cal L}}{\partial\varphi} -
\frac{\partial}{\partial x^{\mu}} \frac{\partial ^{(4)}{\cal
L}}{\partial\varphi_{,\mu}}= F_{\rm ext},
\end{equation}
where $F_{\rm ext} \sim \dot\varphi$ describes a non-conservative
term, which can be related with a Lagrangian interaction. The
origin of this term can be interpreted on the metric (\ref{uu}) as
due to the interaction of $\varphi$ with other fields in in a
thermal bath with temperature $T$.  This term could be the origin
of energy dissipated by the $\varphi$ field into a thermalized
bath.

\section{An example}

To illustrate the earlier results, we can consider the {\em
particular foliation where $\psi^2(t)=3/\Lambda(t)$}. In this case
the effective 4D metric (\ref{u}) is
\begin{equation}\label{uuuu}
\left.dS^2\right|_{\psi(t)=\sqrt{{3\over\Lambda(t)}}}=\left[1 -
\frac{3}{\Lambda(t)}\left(\frac{\dot\Lambda(t)}{2\Lambda(t)}\right)^2\right]dt^2
- \frac{3}{\Lambda(t)} \, e^{2\int \sqrt{\frac{\Lambda(t)}{3}} dt}
\, dr^2,
\end{equation}
so that
\begin{displaymath}
\left.dS^2\right|_{\psi(t)=\sqrt{{3\over\Lambda(t)}}} =
\left.d\sigma^2\right|_{\psi(t)=\sqrt{{3\over\Lambda(t)}}}-
\frac{3}{\Lambda(t)}
\left(\frac{\dot\Lambda(t)}{2\Lambda(t)}\right)^2\, dt^2,
\end{displaymath}
with
\begin{equation}\label{uuu}
\left.d\sigma^2\right|_{\psi(t)=\sqrt{{3\over\Lambda(t)}}}=dt^2 -
\frac{3}{\Lambda(t)} \, e^{2\int \sqrt{\frac{\Lambda(t)}{3}} dt}
\, dr^2\equiv  dt^2 - a^2(t) \, dr^2.
\end{equation}
From the thermodynamical point of view, the hypersurface described
by the brane
$\left.d\sigma^2\right|_{\psi(t)=\sqrt{{3\over\Lambda(t)}}}$ can
be considered as an open system with respect to the closed one
$\left.dS^2\right|_{\psi(t)=\sqrt{{3\over\Lambda(t)}}}$ in
(\ref{uuuu}). We shall consider that (\ref{uuu}) is our physical
spacetime.

The equation of motion for the field $\varphi(t,\vec r)$ on the
metric (\ref{uuu}) [induced from the 5D vacuum defined by the
action (\ref{del}), on the metric (\ref{a1})], is
\begin{equation}
\ddot\varphi + 3 \frac{\dot a}{a} \dot\varphi - \frac{1}{a^2} \,
\nabla^2_r\varphi - \left.\frac{\Lambda(t)}{3}\left[ 4\psi
\frac{\partial\varphi}{\partial\psi} +
\psi^2\frac{\partial^2\varphi}{\partial\psi^2}\right]
\right|_{\psi(t)=\sqrt{\frac{3}{\Lambda(t)}}}=-\frac{\delta}{\dot\varphi},\label{y}
\end{equation}
where the interaction on the right-hand side of (\ref{y}) has its
origin in the temporal dependence of the fifth coordinate on the
brane (\ref{uuu}). In other words, non-conservative terms in the
dynamics of the equation of motion for $\varphi$ are induced by
time dependent terms of $g_{tt}$ in (\ref{uuuu}) omitted in
(\ref{uuu}). For this reason, the interaction $\delta$ is also
dependent on $\dot\psi$, so that
\begin{equation}\label{24}
\delta = 2\,\frac{\dot\psi}{\psi} \dot\varphi^2= \frac{T}{a^3(t)}
\dot{\cal S}
> 0.
\end{equation}
Therefore, for a given temperature $T$, and a given foliation
$\psi(t)=\sqrt{{3\over \Lambda(t)}}$, we can obtain the temporal
evolution of entropy as a function of the fifth coordinate:
\begin{equation}\label{int}
\dot{\cal S}=-\left(\frac{\dot\Lambda(t)}{\Lambda(t)}\right)
\frac{a^3}{T}\, \dot\varphi^2 > 0,
\end{equation}
which means that for $\dot\psi(t) =-
\sqrt{{3\over\Lambda(t)}}\left({\dot\Lambda(t) \over
2\Lambda(t)}\right)>0$, the function $\Lambda(t)$ decreases with
time: $\dot\Lambda(t) <0$. Finally, the eq. (\ref{y}), which
describes the dynamics of $\varphi$ on the FRW metric (\ref{uuu})
from the foliated $\psi(t)=\left(3/\Lambda(t)\right)^{1/2}$
Riemann flat metric (\ref{a1}), is
\begin{equation}\label{movi}
\ddot\varphi + \left[3 \frac{\dot a}{a}
-\frac{\dot\Lambda(t)}{\Lambda(t)}\right] \dot\varphi -
\frac{1}{a^2} \, \nabla^2_r\varphi -
\left.\frac{\Lambda(t)}{3}\left[ 4\psi
\frac{\partial\varphi}{\partial\psi} +
\psi^2\frac{\partial^2\varphi}{\partial\psi^2}\right]\right|_{\psi(t)=\sqrt{\frac{3}{\Lambda(t)}}}=0,
\end{equation}
where the term $\Gamma(t) \dot\varphi=-{\dot\Lambda(t)\over
\Lambda(t)}\dot\varphi$ has a purely dissipative interpretation on
the brane (\ref{uuu}) and describes the energy dissipated by the
inflaton field into a thermalized radiation bath. Notice that the
factor $\left[3 {\dot a\over a} -{\dot\Lambda(t)\over
\Lambda(t)}\right]$ in (\ref{movi}) is the same that whole of the
equation (\ref{dis}). Notice that eqs. (\ref{en}) and (\ref{movi})
are the same equations that describe warm\cite{warm} and
fresh\cite{fre} inflationary scenarios with a Yukawa
self-interaction of $\varphi$: $\delta = \Gamma\,\dot\varphi^2$
and a width $\varphi$-decay: $\Gamma = \left.2{\dot\psi(t)\over
\psi(t)}\right|_{\psi(t)=\sqrt{{3\over\Lambda(t)}}} =
-{\dot\Lambda(t) \over \Lambda(t)}$. In these scenarios,
accelerated expansion and radiation of energy are produced
together during the inflationary stage and the Einstein equations
are
\begin{eqnarray}
H(t)^2 & = &\frac{8\pi G}{3} \rho(t), \\
-(3H(t)^2+2\dot H(t)) & = & 8\pi G\,p(t),
\end{eqnarray}
where $\rho(t)$ and $p(t)$ are respectively given by
\begin{eqnarray}
&& \rho(t) = \frac{\dot\varphi^2(t)}{2} + V(\varphi)+\rho_r(t), \\
&& p(t) = \frac{\dot\varphi^2(t)}{2} -
V(\varphi)+\frac{\rho_r(t)}{3},
\end{eqnarray}
with $V(\varphi) = -\left.{1\over 2} \left({\partial\varphi\over
\partial\psi}\right)^2\right|_{\psi(t)=\sqrt{3/\Lambda(t)}}$ and $H(t)=\dot a(t)/a(t)$ is the Hubble parameter
of the universe on the FRW metric (\ref{uuu}).

Using the fact that the red-shifted temperature evolves as $T(t) =
T_0\,\left[a_0/a(t)\right]$, the eq. (\ref{int}) can be rewritten
as
\begin{equation}\label{ent}
\dot{\cal S} =\frac{\Gamma(t) \, a^4(t)}{T_0\,a_0}\,
\dot\varphi^2\,>0,
\end{equation}
where $T_0$ is the background temperature at $t_0 < t$ and $a_0$
is the scale factor at this moment. This result is a
generalization for 3D spatially flat FRW metrics of whole obtained
using finite temperature quantum field theory by Hosoya and
Sakagami\cite{hos}.

Finally, in order to illustrate our calculations, we can work an
example in which $\Lambda(t) ={3n^2\over t^2}$. In this case
$\Gamma = -{\dot\Lambda(t)\over \Lambda(t)}=2/t$ and the Hubble
parameter is $H(t)={(n+1)\over t}$ for an accelerating universe
with a scale factor $a(t) \sim t^{n+1}$. This implies that
dissipation should be dominant for
$\Gamma(t)/\sqrt{\Lambda(t)\over 3}
>1$\cite{hos}, that is for $n<1$. In other words, dissipation
should be very important in models of inflation where the power of
expansion of the universe
$(n+1)$ is moderated: $1 < (n+1) < 2$. \\

\section{Final Remmarks}

We have shown how dissipative dynamics can be induced (from a 5D
vacuum state) in the inflaton field, which evolves on an effective
4D FRW metric during inflation. A very interesting fact that
results from the analysis is that the entropy increases with time
when $\dot\psi(t) >0$. Furthermore, the expression (\ref{ent})
provide us ${\cal \dot S}(t)$ as a function of geometrical
quantities and the initial temperature. From the point of view of
General Relativity, the emergence of dissipation in the dynamics
of $\varphi$ is a consequence of that the equivalence principle is
broken on the effective 4D FRW metric: $g_{AB} U^A U^B \neq 1$.
This is also the origin of entropy on the FRW metric (\ref{uuuu})
[see eq. (\ref{ent}) for $\dot{\cal S}(t)$, which becomes zero for
$\Gamma=0$], which can be considered as a 4D brane of the 5D
Riemann-flat bulk (\ref{a1}). In this sense, dissipative dynamics
of the inflaton field during inflation in our 4D brane (i.e., our
spacetime) can be considered as indirect evidence of extra
dimensions's existence. However, the system evolves adiabatically
on the metric (\ref{uuuu}), which is comes from a
$\psi(t)=\sqrt{{3\over \Lambda(t)}}$-foliation on the 5D metric
(or bulk) (\ref{a1}).\\

\begin{acknowledgments}
 JR acknowledges UNMdP (Argentina) and MB acknowledges CONICET and UNMdP (Argentina) for financial
support.
\end{acknowledgments}

\end{document}